\documentclass[aps, prd, showpacs, superscriptaddress, ctexart, nofootinbib, twocolumn]{revtex4-1}
\usepackage{graphicx}
\usepackage{amssymb}
\usepackage{color}
\usepackage{enumerate}
\usepackage{overpic}
\usepackage{soul}

\def\beq{\begin{equation}}
\def\eeq{\end{equation}}
\def\beqn{\begin{eqnarray}}
\def\eeqn{\end{eqnarray}}

\renewcommand{\texttt}{{}}
\newcommand{\be}{\begin{eqnarray}}
\newcommand{\ee}{\end{eqnarray}}
\newcommand{\bee}{\begin{equation}}
\newcommand{\eee}{\end{equation}}
\begin{document}

\title{Fermi-bounce cosmology and scale invariant power-spectrum}

\author{Stephon Alexander}
\email{stephon.alexander@dartmouth.edu}
\affiliation{Center for Cosmic Origins and Department of Physics and Astronomy, Dartmouth College, Hanover, New Hampshire 03755, USA}
\author{Cosimo Bambi}
\email{bambi@fudan.edu.cn}
\affiliation{Center for Field Theory and Particle Physics \& Department of Physics, Fudan University, 200433 Shanghai, China}
\author{Antonino Marcian\`o}
\email{marciano@fudan.edu.cn}
\affiliation{Center for Field Theory and Particle Physics \& Department of Physics, Fudan University, 200433 Shanghai, China}
\author{Leonardo Modesto}
\email{lmodesto@fudan.edu.cn}
\affiliation{Center for Field Theory and Particle Physics \& Department of Physics, Fudan University, 200433 Shanghai, China}

\date{\today}

\begin{abstract}
\noindent 
We develop a non-singular bouncing cosmology using a non-trivial coupling of general relativity to fermionic fields. The usual Big Bang singularity is avoided thanks to a negative energy density contribution from the fermions. Our theory is ghost-free since the fermionic operator that generates the bounce is equivalent to torsion, which has no kinetic terms. The physical system consists of standard general relativity plus a topological sector for gravity, and fermionic matter described by Dirac fields with a non-minimal coupling. We show that a scale invariant power-spectrum generated in the contracting phase can be recovered by suitable choices of fermion number density and bare mass, thus providing a possible alternative to the inflationary scenario.  

\end{abstract}


\maketitle

\section{Introduction}

\noindent 
The pioneering work of Hawking and Penrose demonstrated that at initial times, the Friedmann--Lema\^{\i}tre--Robertson--Walker (FLRW) metric of the Standard Big Bang cosmology suffers from singularities in all curvature invariants\cite{HE}. Their theorem states that the initial singularity is unavoidable if space-time is described by General Relativity and if matter obeys the null energy condition. Over the years non-singular bouncing cosmologies have been proposed to avoid the Big Bang singularity by obviating one or all of the assumptions behind the Hawking-Penrose theorem.  However, a successful theory of the early universe must predict the observed nearly scale-invariant spectrum of adiabatic fluctuations in the CMBR.  Scale invariance was attempted in the context of bouncing models with a contracting phase such as Ekpyrotic \cite{Ekpy}, String Gas \cite{stringas} and Pre-Big-Bang scenarios \cite{BF}.  On the other hand, it has proven difficult to obtain adiabatic scale invariant fluctuations in the contracting phase in a number of these models, mainly due to issues in mode matching between the contracting and expanding phases \cite{BF}.  

In pioneering works by Brandenberger, Finelli and independently by Wands \cite{BF,Wands}, it was shown that it is possible to generate a scale-invariant power spectrum in a matter dominated contracting universe.  These authors demonstrated a ``duality'' between the scale invariant power spectrum generated in the inflationary epoch and a contracting matter dominated phase.  During the contracting phase, gauge-invariant perturbations that cross the Hubble-scale are scale-invariant if the scale factor evolves as $a(t) \!\sim\! (-t)^{2/3}$.  Furthermore, if the bounce is non-singular, the scale-invariant modes can be matched to scale-invariant modes in the expanding phase.  

A handful of matter bounce scenarios has since been proposed, mostly based on fundamental scalar fields \cite{quintom, ghost1, ghost2, Ekpy, ek2}. In this paper we present a matter bounce scenario based on Dirac fermions, specifically, on the four-fermion interaction. This line of research has been previously addressed in Ref.~\cite{AP} from the perspective of a wide class of generic potentials of the Dirac field.  Our work differs from Ref.~\cite{AP}, where the four-fermion interaction was neglected in computing the scalar perturbation and a de Sitter background expansion was assumed. 

In this work we follow the perspective taken by ~\cite{AB, Pop} where a torsion induced four-fermion interaction yields a non-singular bounce. In Ref.~\cite{Magueijo:2012ug}, the role of a parity violating 4-fermion self-interaction in a torsion-free theory has been studied. In this work, we propose both a non-minimal coupling of fermions (as analyzed in Refs~\cite{M, BD, BD2}) and a topological gravitational term endowed with torsion (see for instance Ref.~\cite{PR}). We show that these two terms generate four-fermion interactions whose dynamics yield a scale-invariant power spectrum\footnote{The presence of the torsion background turns the topological term in the Holst action from a surface term into a contribution to the four-fermion interaction term.}.

Furthermore, the four-fermion current density modifies the Friedman equations to have a negative energy density that redshifts like $\sim a(t)^{6}$. We show that the resulting bounce is non-singular provided that anisotropic stress is sub-dominant\footnote{In a companion paper \cite{paper00}, some of us used these findings as a starting point to discuss consequences for the fate of black hole solutions \cite{collasso}, which for a  suitable choice of some parameters of the theory may never form.}, consistent with previous literature. Moreover, we show for the first time that the adiabatic quantum fluctuation of fermions in the contracting phase can be scale-invariant. Indeed, as we may straightforwardly infer from the result of Brandenberger and Finelli, since the bounce is non-singular our scale-invariant curvature perturbation induced by the fermion quantum fluctuations will enter the expanding phase as a scale invariant fluctuation.  An advantage of the mechanism shown in this paper is that it does not require any fundamental scalar field. The fermionic field is sufficient to account for both the bounce and the generation of nearly scale-invariant scalar perturbations.  

The following is an outline of the paper. In Section II we introduce the theoretical framework, and cite the relevant works in the literature. In Section III we address the consequences of the model for the Matter-Bounce scenario. In Section IV we address the cosmological perturbations induced by the fermionic field. In Section V we discuss consistency with experimental data. In Section VI we provide some concluding remarks and mention works in progress.

\section{The theory}

\noindent
In what follows we provide our theoretical framework and conventions following the formalism of Refs.~\cite{PR, BD, BD2, AB, literature, ABC}. We start by considering a generalization of the Einstein-Hilbert action with a topological term: this is the Holst action for gravity in the Palatini formalism which allows us to couple gravity to chiral fermions. We then couple this theory to a Dirac field $\psi$, whose complex conjugate reads $\overline{\psi}=\left(\psi^{*}\right)^{T}\gamma^{0}$. The action for the fermionic field is cast in terms of the Dirac matrices, $\gamma^I$ with $I=0,\ldots,3$ and $\gamma^5$, expressed in the Dirac-Pauli basis. The action for pure gravity can be cast in terms of the gravitational field $g_{\mu \nu}=e_{\mu}^I e_{\nu}^J \eta_{IJ} $, where $e^I_\mu$ is the tetrad/frame field (with inverse $e_I^\mu$ and determinant $e$), and the Lorentz connection $\omega^{IJ}_\mu$ (whose curvature is $F^{I J}_{\mu\nu}=2 \partial_{[\mu}\omega^{IJ}_{\,\nu]} + \left[ \omega_\mu, \omega_\nu \right]^{IJ}$). The action for the fermion fields involves the spinors $\psi$ and $\overline{\psi}=\psi^\dagger \, \gamma^0$. 

The total action is the sum of the Einstein-Cartan-Holst (ECH) action plus the non-minimal covariant Dirac action \footnote{Notice that, in absence of the gravitational Holst topological term, the whole theory provided with torsion and minimally-coupled fermions is referred to in the literature as the Einstein-Cartan-Sciama-Kibble theory. See {\it e.g.} Refs.~\cite{literature}.}.  The ECH action is  (see \cite{paper00}), 

\begin{eqnarray}
\label{nonminimalaction1}
&&{S}_{\rm Holst} =
\frac{1}{2 \kappa} \int_{M}\!\!  d^{4}x \;|e| \, e^{\mu}_{I}e^{\nu}_{J}
P^{IJ}_{\ \ \ KL}F^{\ \ KL}_{\mu \nu}(\omega)\, , 
\end{eqnarray}
where $\kappa = 8 \pi G_{\rm N}$ is the reduced Planck length square and the operator $P^{IJ}_{\ \ \ KL}=\delta^{[I}_{K} \delta^{J]}_{L} - \epsilon^{IJ}_{\ \ KL}/ (2 \gamma)$, $\epsilon_{IJKL}$ being the Levi-Civita symbol, is defined in terms of the Barbero--Immirzi parameter $\gamma$, and can be inverted for $\gamma^2\neq -1$. 
The Dirac action is $S_{\rm Dirac} = \!\frac{1}{2 }  \int d^{4}x |e| \mathcal{L}_{\rm Dirac}$, where
\begin{eqnarray}
\label{nonminimalaction2}
\mathcal{L}_{\rm Dirac} = \!\frac{1}{2 } \!\left[\overline{\psi}\gamma^{I}e^{\mu}_{I}\!\left(1-\frac{\imath}{\alpha}\gamma_{5}\right)\!\imath \nabla_{\mu}\psi -  m \overline{\psi} \psi \right] + {\rm h.c.}
\, , 
\end{eqnarray}
in which $\alpha\in \mathbb{R}$ is the so called non-minimal coupling parameter. The Einstein-Cartan action can be found if we consider $S_{\rm ECH}\!\!=\!\!S_{\rm GR} \!+\! S_{\rm Dirac}$ and $\alpha\!=\!\gamma$, with a term that reduces to the Nieh-Yan invariant ~\cite{M} 
when the second Cartan structure equation holds. From the point of view of the Holst action (\ref{nonminimalaction1}), minimal coupling is recovered in the limit $\alpha \rightarrow \pm\infty$. Constraints on $\alpha$ and $\gamma$ can be derived from the four fermion axial-current Lagrangian (\ref{interact}),  based on measurements of lepton-quark contact interactions \cite{Freidel:2005sn, deBlas:2013qqa}, but these are not at all stringent.  

The covariant derivative for Dirac spinors is defined to be $ \nabla_{\mu}\equiv \partial_{\mu} + \frac{1}{4}\omega^{IJ}_{\mu} \gamma_{[I} \gamma_{J]}$, while the field-strength of the Lorentz connection is obtained from $\left[\nabla_{\mu},\nabla_{\nu}\right] = \frac{1}{4}F^{IJ}_{\mu \nu}\gamma_{[I} \gamma_{J]}$. 
Because of the presence of fermions, a torsional part of the connection enters the non-minimal ECH action. Nevertheless, the latter can be integrated out of the theory through the Cartan equation, which is found by varying the total action with respect to the connection $\omega^{IJ}_\mu$. We provide the usual definition of the contortion tensor, denoted as $C_\mu^{IJ}$ and defined by $(\nabla_\mu-\widetilde{\nabla}_\mu ) V_I = C_{\mu \,I}^{\ \ J}\, V_J$, where $\widetilde{\nabla}_\mu$ is the covariant derivative compatible with the tetrad $e^I_\mu$ and $V_J$ a vector in the internal space.  The Cartan equation then relates the contortion tensor $C_\mu^{IJ}$ to the fermionic currents and tetrad:
\be \label{carta}
&& e^\mu_I \, C_{\mu JK} = \frac{\kappa}{4} \, \frac{\gamma}{\gamma^2 +1} \, \left( \beta \, \epsilon _{IJKL}\ J^L - 2 \theta \, \eta_{I[J} \, J_{K]}  \right) \,, \nonumber \\
&&J^L=\overline{\psi} \gamma^L \gamma_5 \psi 
\, ,
\ee
where the coefficients are functions of the free parameters within the non-minimal ECH theory, $\beta=\gamma+1/\alpha$ and $\theta=1-\gamma/\alpha$. Thanks to (\ref{carta}) the non-minimal ECH action can be completely recast in terms of the metric compatible connection, as a sum of the Einstein-Hilbert action and the Dirac action. The latter is now written in terms of metric compatible variables, and now includes a novel interaction term that captures the new physics within the non-minimal ECH theory $S_{\rm ECH}$.   The theory then becomes:
\be
S_{\rm ECH}= S_{\rm GR} + S_{\rm Dirac}
+S_{\rm Int}\,,
\ee
where the Einstein-Hilbert action is expressed in terms of the mixed-indices Riemann tensor $R_{\mu\nu}^{IJ}=F_{\mu\nu}^{IJ}[\widetilde{\omega}(e)]$
\be
S_{GR}= \frac{1}{2 \kappa} \int_{M} \!\!\! d^4 x |e| e^\mu_I e^\nu_J R_{\mu\nu}^{IJ} \,, 
\ee
the Dirac action $S_{\rm Dirac}$ on curved space-time 
reads
\be
S_{\rm Dirac}= \frac{1}{2} \int_{M} \!\!\! d^4 x |e| \left( \ 
\overline{\psi} \gamma^I e^\mu_I \imath \widetilde{\nabla}_\mu \psi - m \overline{\psi} \psi \right) +{\rm h.c.}\,,
\ee
and the interacting term is:
\be \label{interact}
S_{\rm Int} \!=\! -\xi \kappa\! \int_{M} \!\!\! d^4 x |e| \,J^L\, J^M\, \eta_{LM}\,,
\ee
where we define the coefficient $\xi$ as a function of the fundamental parameters of the theory,
\be
\xi:= \frac{3}{16} \!\frac{\gamma^2}{\gamma^2+1}\! \left(1 + \frac{2}{\alpha \gamma} -  \frac{1}{\alpha^2} \right)\,.
\ee 

In what will follow it is useful to compute the energy-momentum tensor,

\be \label{tenfe}
\hspace{-0.25cm}
T^{\rm fer}_{\mu\nu}\!=\! \frac{1}{4}  \overline{\psi} \gamma_I e^I_{( \mu} \imath \widetilde{\nabla}_{\nu )} \psi +{\rm h.c.}  -
 g_{\mu\nu} \mathcal{L}_{\rm fer}.
\ee  

In canonical quantum field theory, spinors are operator-valued fields $\hat{\psi}$ that act on a definite Hilbert space. We can express a classical spinor as the expectation value of the spinor operator on an appropriate quantum state $| s\rangle$, such that $\psi=\langle s | \hat{\psi} |s \rangle$, which is a complex number. The observable bilinear that will enter the classical equation will be evaluated on such a quantum state, and their renormalized value will be obtained by subtracting the vacuum expectation value, namely $\langle \dots \rangle_{\rm ren}\equiv \langle s| \dots |s \rangle - \langle 0| \dots |0 \rangle $.

The Dirac equation on a curved background for the interacting system is found to be\footnote{To further simplify the four-fermion term, we have used the Pauli-Fierz identity
\begin{eqnarray}
(\overline{\psi} \gamma_5 \gamma^I \psi)  (\overline{\psi} \gamma_5 \gamma_I  \psi)
= (\overline{\psi}  \psi)^2 + (\overline{\psi} \gamma_5 \psi)^2 + (\overline{\psi} \gamma^I  \psi)  (\overline{\psi}  \gamma_I  \psi)  \, .
\nonumber
\end{eqnarray}
}

\be
\hspace{-0.2cm}
\!\!\!\!\!\gamma^I \!  e_I^\mu \imath \widetilde{\nabla}_{\mu} \psi - m \psi = 2 \xi \kappa  (\overline{\psi} \psi +\overline{\psi} \gamma_5 \psi   \gamma_5 + \overline{\psi} \gamma_I  \psi  \gamma^I) \psi .
\ee   

\section{Non-Singular Bounce}

\noindent
In previous bouncing models, the issue of the robustness of the singularity avoidance depends on whether quantum corrections ({\it i.e.} curvature or matter) were under-control at the bounce \cite{BF}.  The advantage of our model is that torsion in this scheme, which is responsible for the bounce, has no kinetic term ({\it i.e.} it is an auxiliary field) and will not experience any quantum corrections as we approach the bounce.  

We would like to find self-consistent initial values of the fermionic densities so as to not spoil isotropy of our FLRW space-time.  First we cast our metric $e^I_\mu$ in FLRW form, which in the comoving gauge reads $e^I_0=\delta^I_0$ and $e^I_j=\delta^I_j \, a(t)$. Homogeneity and isotropy on spatial hyper-surfaces demand a vanishing fermionic current. 

As for the Dirac field components, the vanishing of the spatial fermionic current yields \cite{AP} $\psi\!=\!(\psi_0,\, 0\,,0\,,0)$. In the  comoving gauge, the only non-vanishing spin connection components for $\omega^{IJK}\!=\!\omega^{IJ}_\mu\, e^\mu_K$ are $\omega_{0ij}=-\omega_{i0j}= - H \delta_{ij}$, where the Hubble parameter is defined by $H=\dot{a}/a$ and ``$\dot{\phantom{a}}$'' represents the time-derivative. This implies $\widetilde{\nabla}_0\!=\!\partial_0$ and $\widetilde{\nabla}_i\!=\!\partial_i+ a H/2 \delta_{ij}{\rm diag} (\sigma^j, - \sigma^j)$, where $\sigma^j$ denotes Pauli matrices. The Dirac equation then follows
\be
\dot{\psi}_0 \!+\! \frac{3}{2} \,H\, \psi_0 \!+\!\imath \left(m\!+\! 4\kappa\, \xi \,  \psi_0^\star \psi_0 \right) \!\psi_0\!=\!0,
\ee
in which $^\star$ denotes complex conjugation. It is easy to derive the equation of motion for the bilinear $\psi_0^\star\psi_0$, 
\be \label{seque}
\frac{d }{dt} \, \psi_0^\star \psi_0+ 3\,H\, \psi_0^\star \psi_0 =0\,,
\ee
which yields the familiar expression in terms of a constant initial density $n_0$
\be \label{ro}
\psi_0^\star \psi_0\sim \frac{n_0}{a^3} \,.
\ee

Using the solutions of (\ref{seque}) the Friedmann equation becomes,
\be \label{H2}
H^2 =\xi\,\frac{\kappa^2}{3}\,\frac{n_0^2}{a^6} + \frac{m \,\kappa}{3}\frac{n_0}{a^3} \,.
\ee 

We see from (\ref{H2}) that the four-fermion term has the crucial $\frac{1}{a^6}$ redshifting which will control the bounce.  We now consider a contracting scale factor and immediately recognize that the bounce is due to the vanishing of the total energy density.  As we approach the would be singularity (the scale-factor approaching zero), the negative energy (for $\xi<0$) four-fermion term dominates and drives the Hubble parameter to zero, resulting in a non-singular bounce.

At the bounce we will have to obtain the initial value of $\dot{H}$ which we can get from the second Friedmann equation,

\be\label{Hpunto}
\dot{H}-H^2=\frac{\ddot{a}}{a}= -  \frac{1}{6} \left(\! m \kappa\, \frac{n_0}{a^3} + 4 \,\xi \kappa^2\, \frac{n_0^2}{a^6}  \right)\!.
\ee 

At the bounce, $t=t_0$, the vanishing of $H=H_0$ in (\ref{H2}), the scale factor approaches a constant, $a_0=(-\xi \kappa n_0/m)^{1/3}$. For negative values of the $\xi$ parameter, the scale factor $a_0$ reaches its minimum, as from (\ref{Hpunto}) one finds that $\dot{H}_0=- m^2/(3 \xi)$. Notice that both the bilinear $\overline{\psi} \psi$ and the field $\psi$ reach their maxima at $t_0$: although the effective potential in $S_{\rm fer}$ is unbounded in $\psi$ when $\xi$ is negative, the gravitational bounce prevents the classically unbounded energy spectrum from taking infinite negative values. It is then straightforward to find the deterministic evolution of the scale factor that leads to the bounce:
\be \label{adri}
a=\left( \frac{3 m \kappa n_0}{4} (t-t_0)^2 - \xi \frac{\kappa n_0 }{m}  \right)^{\frac{1}{3}} \,.
\ee
This solution can be shown to be stable under perturbations to the fermionic matter field if  the anisotropic and inhomogeneous contribution to the energy density, which reads
\be
\tilde{\rho} \sim \frac{ \rm{Tr}{[\gamma_i \gamma_j]} }{M_p^2} \, \overline{\psi} \psi \, \langle \delta \overline{\psi} \delta \psi \rangle
\,,
\ee
is subdominant with respect to the isotropic contribution in the right hand side of (\ref{H2}). The criterion to have a subdominant contribution is 
\be
 \langle \delta \overline{\psi} \delta \psi \rangle/M_p^2 <\!\!< m\,. 
\ee

We show in Fig.~1 the range of values in the ($\gamma$,\,$\alpha$) parameter-space for which $\xi$ is negative, and the matter-bounce happens. The bounce takes place when the interaction energy of the fermion fields provide a negative contribution that violates the null energy condition, as shown for non-conventional fermion fields in Refs.~\cite{AP, Alexander:2008vt, Cai:2013rna}. 
\begin{figure*}[t]
\begin{center}
\hspace{1cm}
\begin{overpic}[scale=0.625]{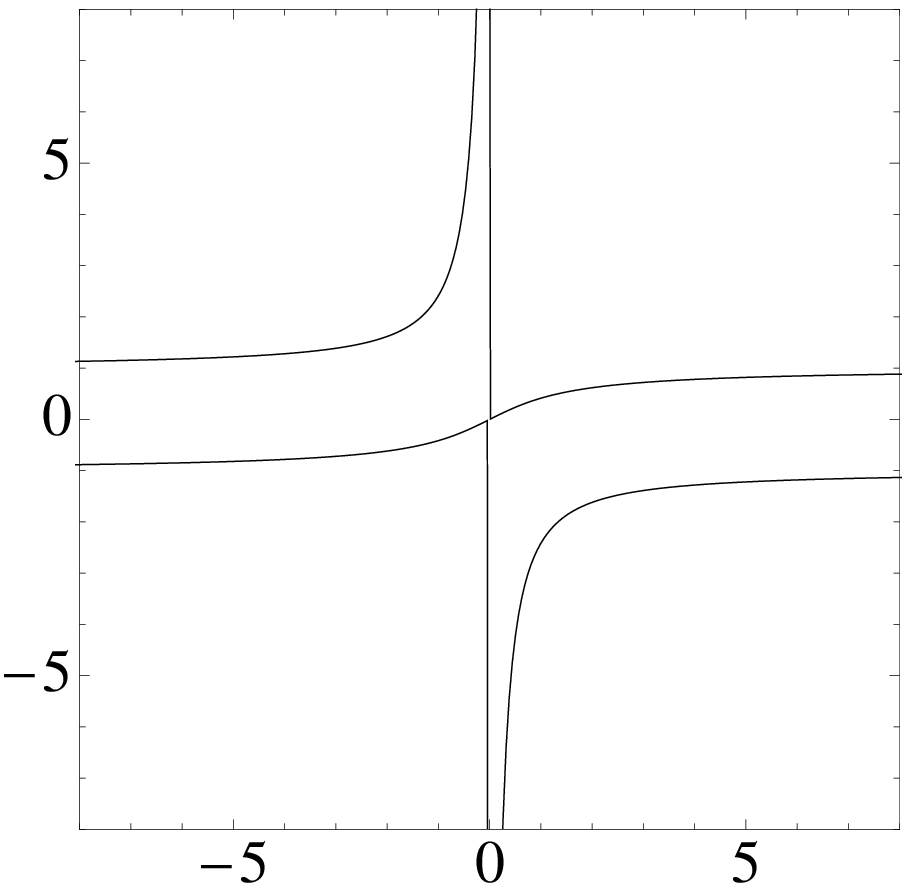}
\put(-6,50){$\alpha$}
\put(50,-6){$\gamma$}
\put(25,75){$\xi>0$}
\put(25,53){$\xi<0$}
\put(25,25){$\xi>0$}
\put(70,75){$\xi>0$}
\put(70,50){$\xi<0$}
\put(70,25){$\xi>0$}
\end{overpic}
\end{center}
\vspace{0.5cm}
\caption{Sign of $\xi$ on the plane ($\gamma$,\,$\alpha$). Natural values of the fundamental parameters are allowed in order to obtain $\xi<0$.} 
\label{fig} 
\end{figure*}

\section{Cosmological Curvature Perturbations} 

\noindent
In this section, we demonstrate that fermions can induce scale-invariant adiabatic perturbations. We show this in a way similar to the treatment of scalar fields, in that we solve the mode functions for the fermionic perturbations and evaluate its contribution to the gauge invariant curvature perturbation. Following \cite{AP}, we introduce a quantity that is conserved on large scales and can be related to CMBR temperature fluctuations, analogous to the Bardeen variable\footnote{The cosmological models whose parameters $\alpha$ and $\gamma$ encode a negative $\xi$ may have application as alternative model of Inflation. Below we study the perturbation variable sourced by fermionic matter.}, {\it i.e.}
\be
\zeta=\frac{\delta \rho}{\rho+p}. 
\ee
After some algebra and the use of the Pauli-Fierz identity, we obtain
\be 
\!\!&& \zeta = \frac{1}{ m \overline{\psi} \psi + 2 \xi \kappa J_L J^L} \Big\{
\left( m + \xi \kappa  \overline{\psi} \psi \right)  \left(\delta\overline{\psi} \, \psi+ \overline{\psi} \, \delta \psi \right)+ \nonumber\\
&&\,\, \xi \kappa \Big[ \overline{\psi} \gamma_5 \psi ( \delta\overline{\psi} \, \gamma_5 \psi +\overline{\psi} \gamma_5 \delta \psi ) \!+\! \overline{\psi} \gamma^L \psi ( \delta\overline{\psi} \gamma_L \psi + \overline{\psi} \gamma_L \delta \psi )  \Big] \!\Big\} .  \nonumber
\ee
We can further simplify $\zeta$ by using the background solutions for the spinors:
\be 
\zeta &=& f(t)(\delta\overline{\psi} \psi +  \overline{\psi} \delta \psi)\,,\nonumber \\
{\rm with} \,\,\,  f(t) &\simeq& \frac{(1 - \xi \kappa \overline{\psi} \psi/m )}{\overline{\psi} \psi}\,.
\ee

The power spectrum $\mathcal{P}(k)$ is implicitly  defined in terms of the equal-time two-point function of $\zeta$:
\be
\langle \zeta(t, \vec{x}) \zeta(t, \vec{x}+\vec{r}) \rangle = \int \frac{dk} {k} \, \frac{\sin k r}{kr}\, \mathcal{P}(k)
\,,
\ee
where the expectation value is taken in the vacuum state and defined by $a |0\rangle = b |0\rangle = 0$. Using the simplified expression for $\zeta$ in terms of $\delta\psi$ and the background Dirac field, the two point function becomes

\be
\langle  \zeta(t, \vec{x}) \zeta(t, \vec{x}) \rangle = f^2(t) \frac{\overline{\psi} \psi}{4}  \langle \delta \overline{\psi}\, \delta\psi \rangle \,.
\ee
This finally provides the expression for the power-spectrum
\be \label{PS}
\hspace{-0.3cm}
\mathcal{P}(k)\sim \sum_h\,\frac{m a^3(t) -2 \xi \kappa n_0 }{4 m n_0} \ \frac{k^3}{4 \pi^2} \,  \overline{v}_h (t, \vec{k}) \, {v}_h (t, \vec{k})\,,
\ee
where the mode function $v_h$ is obtained by expanding the quantum fluctuations of the spinor:
\be
&& \hspace{-0.4cm}
\delta \psi = \sum_h \int \!\! \frac{d^3k}{(2 \pi)^{3/2}}  \times \\
&& \hspace{0.6cm}
 \Big( u_h{(t, \vec{k})} a{(\vec{k}, h)} e^{\imath \vec{k} \cdot \vec{x}} + v_h{(t,\vec{k})} b^\dagger{(\vec{k}, h)} e^{-\imath \vec{k} \cdot \vec{x}} \Big) .\nonumber
\ee 

We now proceed to evaluate the solution of the mode function $v_h$ of the spinor perturbation. Using the background solution of $\psi$, and a conformal rescaling, $\widetilde{\psi}=a^{3/2} \psi$, we obtain the equation of motion for the spinor perturbation

\be \label{Diri}
\left( \imath \gamma^\mu \partial_\mu - m \, a(\eta) -\frac{2 \xi \kappa \, n_0}{a^2(\eta)} \right) \widetilde{\delta\psi}=0 \, .
\ee 

We can now solve the Dirac equation (\ref{Diri}) in terms of the following mode\footnote{Given a unit eigenspinor $\xi_{h}$, the helical components of the mode functions are

\begin{eqnarray} 
&& \hspace{-0.8cm} \nonumber
\tilde{u}(t, \vec{k}) = \sum_h \tilde{u}_h(t, \vec{k})= \sum_h \left( 
\begin{array}{c} \tilde{u}_{L, h} (\vec{k}, \eta) \\ \tilde{u}_{R,h} (\vec{k}, \eta)  \end{array}\right) \xi_h\,, \\
&& \hspace{-0.8cm}  \nonumber
\tilde{v}(t, \vec{k}) = \sum_h \tilde{v}_h(t, \vec{k})= \sum_h \left( 
\begin{array}{c} \tilde{v}_{R, h} (\vec{k}, \eta) \\ \tilde{v}_{L,h} (\vec{k}, \eta)  \end{array}\right) \xi_h\,.
\end{eqnarray}}
functions:

\be
&& \tilde{f}_{\pm h}= \frac{1}{\sqrt{2}} [\tilde{u}_{L,h} (\vec{k}, \eta)+\tilde{u}_{R,h} (\vec{k}, \eta)]\,, \nonumber\\
&&  \tilde{g}_{\pm h}=  \frac{1}{\sqrt{2}} [\tilde{v}_{L,h} (\vec{k}, \eta)+\tilde{v}_{R,h} (\vec{k}, \eta)]\, . 
\ee
Equation(\ref{Diri}) can be expressed in terms of $f_{\pm h}$:
\be 
&& 
\tilde{f}''_{\pm h} + \left[k^2 \!+\! m^2 a^2+ \imath m a' 
+ 2 \xi \kappa n_0 \!\left(\frac{m}{a} \!-\! \imath \frac{a'}{a^3}\right) \right] \tilde{f}_{\pm h}=0, \nonumber \\
&& \label{spinors}
\ee
where $\pm h$ denotes helicity and the and $'$ denotes derivative with respect to conformal time.  An identical system of coupled equations is recovered for $\tilde{g}_h$. Rescaling (\ref{spinors}) by $\kappa$ and then taking the limits $\eta_0^2<\!\!<\kappa$ and $\kappa m^2<\!\!<1$, provided that also $\kappa^2 m^2\!<\!\!<\! \eta_0^2$ is fulfilled, equation (\ref{spinors}) reduces to
\be \label{ino}
\!\!\tilde{f}''_{\pm h} \!\!+\!\! \left(k^2 \!-\! \frac{\nu^2-1}{4\, \eta^2} \! \right) \!\tilde{f}_{\pm h}\!=\!0\,,
\ee
where the parameter $\nu$ is related to the fermion coupling parameter $\xi$ by
\be \label{nuxi}
\nu^2 = 1 - 8 \xi\,.
\ee 
An equation identical to (\ref{ino}) is then found for $\tilde{g}_{\pm h}$; their solutions have been extensively studied in the literature, and for non densitized components read 
\be \label{effe}
f_{\pm h}(k, \eta) = \sqrt{ \frac{m}{k}} \,\sqrt{\frac{-\pi k \eta}{8 a^3(\eta)}}\, Z_{|\nu|}(- k \eta)\,,
\ee
$Z_{|\nu|}$ denoting the Bessel functions labeled by the parameter $|\nu|$.  In the contracting epoch and on sub-horizon scales, when $-k \eta\!>\!\!>\!1$, both $f_{\pm h}(k, \eta)$ and $g_{\pm h}(k, \eta)$ the fermionic perturbations are oscillatory and suppressed by a factor $a^{3/2}(\eta)$. In this limit, the factor $\sqrt{m/k}$ in (\ref{effe}) determines the quantum vacuum initial conditions \cite{robra}. 

 For super-Hubble perturbations, {\it i.e.} $-k \eta\!<\!\!<\!1$, the solutions of (\ref{ino}) are Bessel functions, $Z_{|\nu|}\simeq \Gamma(|\nu|) (-k \eta)^{-|\nu|/2-1/2}$, in which $\Gamma(|\nu|)$ denotes the Euler function\footnote{ For any value of $\nu$, this provides perturbations $\delta\psi(\eta, \vec{x})$ and $\delta\psi'(\eta, \vec{x})$ which decrease during the expanding phase of the universe.}. We now see that $\nu$, which is related to four-fermion coupling parameter, determines a scale invariant power-spectrum.   Far away from the bounce, the power-spectrum (\ref{PS}) can be evaluated to be
\be \label{PK}
\mathcal{P}(k) \simeq  \frac{m\,k^2 |\Gamma(|\nu|)|^2}{16\, n_0} \, | k \eta|^{- |\nu|} \nonumber \,.
\ee
A value of $|\nu|=2$ then ensures scale-invariance, providing the expression for the power-spectrum 
\be \label{Ps}
\mathcal{P}(k)\simeq  \frac{m}{16\, n_0 \eta^2}\,,
\ee
which is evaluated at the end of the matter contracting phase $t_E$, at which the scale factor takes the value $a_E$. The resulting expression would then be
\begin{eqnarray}
\label{P_S_final}
\mathcal{P}_S \simeq  \frac{m H_E^2}{32\, n_0 } \,,
\end{eqnarray}
where $\eta_E = 2/(a_E {\cal H}_E)=2/ H_E$ has been applied. Notice that the time $t_E$ marks the moment at which perturbations become constant, throughout the rest of the primordial epoch, until they reenter the Hubble radius.

\section{Consistency with observations}

\noindent
An exact scale-invariance of the power spectrum would immediately constrain the parameter $\xi$ to take the value
\be
|\xi|= \frac{3}{8} \,.
\ee
But observed deviations from scale invariance, namely $n_s=0.960\pm0.007$ \cite{Ade:2013zuv}, as parametrized from (\ref{PK}) through the relation 
\be
n_s-1\equiv \frac{d \ln \mathcal{P}(k)}{d \ln k} = 2- |\nu|
\,,
\ee
requires a slightly different value for $\xi$, {\it i.e.}
\be \label{mars}
\xi = \frac{1-(3-n_s)^2}{8}\simeq - 0.395\pm 0.004\,,
\ee
once we have taken into account (\ref{nuxi}).

Notice that the value of $\xi$ consistent with the CMB (\ref{mars}) will restrict the bare parameters in our theory ($\gamma$ and $\alpha$) to one-parameter family of theories.  Finally, the choice $|\xi|\simeq 4 \cdot 10^{-1}$ is also clearly consistent with particle physics data, given the lack of a stringent constraint coming from the lepton-quark contact interactions. Measurements constrain $|\xi|\!<\! 10^{32}$ \cite{Freidel:2005sn, deBlas:2013qqa}, which in turn may allow a region of natural values for the parameters entering the non-minimal Einstein-Cartan-Hilbert theory resulting from (\ref{nonminimalaction1}) and (\ref{nonminimalaction2}). 

Recently, there has been much discussion about the possible detection of primordial gravitational-waves by the BICEP2 collaboration \cite{Ade:2014xna}. The result has since been questioned in the literature by a few studies (see {\it e.g.} \cite{Mortonson:2014bja} and \cite{Flauger:2014qra}), which point out possible flaws in the data analysis. It has been shown that a proper dust profile might still account for all or most of the signal of the primordial gravitational waves \cite{Ade:2013zuv}.

With a view towards more detailed data analyses to be delivered by BICEP2 and other collaborations, it is sensible in this work to show the derivation of the phenomenological parameter $r$, which accounts for the ratio between the primordial gravitational waves' power-spectrum and the scalar perturbations power-spectrum. This can be achieved, recalling that at the perturbative level both the scalar and the tensor metric fluctuations can be treated linearly and as uncoupled degrees of freedom. Thus the derivation of the primordial gravitational waves' power-spectrum will be immediately achieved following the standard procedure outlined in \cite{TCP}, which is specialized to general matter-bounce scenarios (see {\it e.g.} \cite{Cai:2013kja}). The primordial gravitational power-spectrum is:
\begin{eqnarray} \label{PTT}
 \mathcal{P}_T = \frac{1}{\vartheta^2}\frac{H_E^2}{ M_p^2}\,,
\end{eqnarray}
where $\vartheta = 8\pi(2q-3)(1-3q)$ (the coefficient $q$ is a background parameter associated with the contracting phase and typically required to be less than unity), and the comoving Hubble parameter $H_E$ is evaluated at the end of matter contracting phase, just before the phase transition to the bounce. The maximal amplitude of the Hubble rate can then be evaluated from requiring the scale factor to be of order $\frac{m}{\sqrt{\xi}}$, once the Universe has been assumed to evolve through the bounce. 
 
Most of the bouncing models that generate adiabatic fluctuations in the contraction phase (before the bounce), including the Ekpyrotic scenario, would be disfavored or eventually ruled out if the claim by the BICEP2 collaboration \cite{Ade:2014xna} on the detection of B-modes coming from primordial gravitational waves, and the related value of the tensor to scalar ratio $r\simeq0.2$, is confirmed. The model in this paper, consisting of only one fermionic species, would then suffer a similar fate, as the theoretical value for $r$ consistent with the mass parameter of the model is found to be too large. Indeed, it follows from (\ref{P_S_final}) and (\ref{PTT}) that
\begin{eqnarray}
r\simeq \frac{32}{\vartheta^2}\, \frac{n_0}{m M_p^2}\,,
\end{eqnarray}
which can not match experimental constraints consistently with the conditions $\eta_0^2\!\!<\!\!<\! \kappa$ and $\kappa m^2 \!\!<\!\!<\!1$ previously required for scale-invariance.  Therefore we conclude that if the BICEP2 detection is confirmed in the future, then our specific model could be ruled out\footnote{Nevertheless, the introduction of a second fermionic species in the analysis can account for a new degree of freedom able to match a smaller value of $r$. This latter argument has been explored in \cite{ultimo}. }.  
\section{Summary and conclusion}  

\noindent
When general covariance accommodates non-minimal coupling in the fermionic sector, a four-fermion interaction modifies the cosmological evolution to yield a bounce.  In this work we have demonstrated that the same fermions that regulate the singularity also generate scale-invariant quantum fluctuations in the contracting phase. Using the arguments of Brandenberger and Finelli, we can easily match these fermionic perturbations to the scale invariant modes in the expanding phase.  The bounce is non-singular because the torsion, which is responsible for the bounce, does not receive quantum corrections. The gravitational wave power spectrum and the resulting tensor to scalar ratio have been derived. In a future paper, it will be interesting to compute corrections to the tensor to scalar ratio due to the coupling of the gravitons to the fermions. 

Furthermore, in order to fully understand the generation of scale-invariant scalar perturbations, it is essential to address the mechanism here discussed in terms of the canonical Mukhanov-Sasaki variables, to which both matter and metric perturbations contribute. In a forthcoming paper \cite{nuova} some of us are considering how to recover those variables by looking at the second order action of the theory.  In this context, it is interesting to notice that while matter perturbations are derived from the relevant fermionic bilinears, which behave as scalar and vector fields, perturbations of the fermionic bilinears must be expressed in terms of the fundamental fermionic fields, the dynamics of which is dictated by first order differential equations. This feature is at the origin of the very different behavior of fermionic matter perturbations relative to scalar field perturbations.

\begin{acknowledgments}
\noindent
We thank the Referees for their valuable remarks and suggestions. We dedicate this paper to Leon Cooper, whose work continues to inspire and challenge us. SA was supported by the Department of Energy Grant DE-SC0010386. This work was supported by the NSFC grant No.~11305038, the Shanghai Municipal Education Commission grant for Innovative Programs No.~14ZZ001, the Thousand Young Talents Program, and Fudan University. 
\end{acknowledgments}

\end{document}